05.00.00 Technical sciences

05.00.00 Технические науки

UDC 51-74

# On the Variety of Planar Spirals and Their Applications in Computer Aided Design

[1] Rushan Ziatdinov
[2] Kenjiro T. Miura

[1] Department of Computer & Instructional Technologies,
Fatih University, 34500 Büyükçekmece, Istanbul, Turkey
PhD (Mathematical Modelling), Assistant Professor
E-mail: rushanziatdinov@gmail.com, ziatdinov@fatih.edu.tr
URL : http://www.ziatdinov-lab.com/
[2] Department of Information Science and Technology,
Graduate School of Science and Technology, Shizuoka University,
3-5-1, Johoku, Naka-ku, Hamamatsu Shizuoka, 432 Japan
PhD (Mechanical Engineering), Professor
E-mail: tmkmiur@ipc.shizuoka.ac.jp

**ABSTRACT.** In this paper we discuss the variety of planar spiral segments and their applications in objects in both the real and artificial world. The discussed curves with monotonic curvature function are well-known in geometric modelling and computer aided geometric design as fair curves, and they are very significant in aesthetic shape modelling. Fair curve segments are used for two-point $G^1$ and $G^2$ Hermite interpolation, as well as for generating aesthetic splines.

> "The human mind always makes progress, but it is a progress in spirals."
> - Madame de Staël

**1. INTRODUCTION.** Historically spirals have been known as positive symbols representing eternity, growth and evolution, and they have been utilized for thousands of years. For the ancient Greeks, the spiral represented the symbol of infinity, constant motion and awareness. For the Celts, the spiral represented growth, birth and expansion (Fig. 1). The African spiral, dwennimmen shown on Fig. 2, is a symbol of strength and humility. In many ancient cultures the spiral was associated with the path leading from outer consciousness to the inner soul.

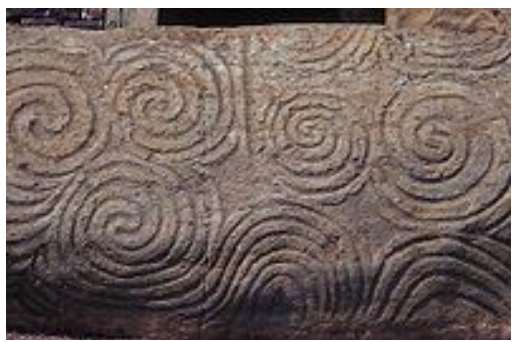

Figure 1: The triple spiral or triskele is a Celtic and pre-Celtic symbol.

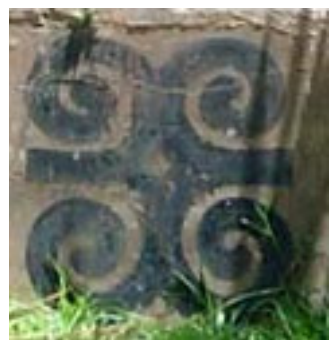

Figure 2: A **dwennimmen**. It is a symbol of humility and strength found in a town called Aburi, inside the botanical gardens.

In addition to ancient artificial paintings, spirals are known to be an inherent part of the natural world: horns, seashells, bones, leaves, flowers and tree trunks (Cook, [1]; Harary and Tal, [2]). Nowadays many well-known spirals are used as attractive curves in computer-aided design and styling, and have wide usage from jewellery design (earrings, pendants) [3] to aircraft design.

**2. Classification of planar spirals**

There is a huge variety of spiral curves which can be classified into three groups:

1. Algebraic spirals

These are spirals whose equations in polar coordinates are algebraic with respect to the variables $\rho$ and $\phi$. These include the hyperbolic spiral; the Archimedean spiral; the Galilean spiral; the Fermat spiral; the parabolic spiral; the lituus; the atom* spiral [18]; the Pritch-Atzema spiral [18] which is the curve whose catacaustic is a circle; cochleoid [18]; the Cotes' spiral which can be defined as the solution to the central orbit problem [18], etc. Their possible applications in computer-aided geometric design (CAGD) still remain unknown.

2. Fair curves

These are also known as curves with a monotonic curvature function. They include Class A Bézier curves [4], Pythagorean-hodograph curves of monotone curvature [5], log-aesthetic curves (LACs) [6-10], formerly known as pseudospirals [11], as well as generalized log-aesthetic curves (GLACs) studied in [12-13]. LACs and GLACs are generalizations of the well-known Cornu, Nielsen and logarithmic spirals, and involutes of a circle. Superspirals introduced in [14] are the most general fair curves, whose equations are given in terms of the Gaussian hypergeometric function. Fair curves, as well as fair surfaces, are used by designers for generating aesthetically pleasing shapes and are very important in product lifecycles. Harada's quantitative analysis method of the characteristics of various curves used in artificial objects [6] has shown that many high-quality curves have linear logarithmic curvature and torsion graphs [15].

3. Discrete spirals

Discrete spirals are represented by points and linear segments, which connect corresponding points. One of the most famous discrete spirals is the Theodorus spiral, or square root spiral [16-17], which is constructed from a sequence of right triangles with sides $(\sqrt{n}, 1, \sqrt{n+1}), n = 1, 2...$ (Fig. 3). The other discrete spirals are the golden spiral (Fig. 4), the Pinwheel spiral [16] and the spirangle (Figs. 5 and 6) useful in ophthalmology as vectograms and in electronics as printed inductors, etc.

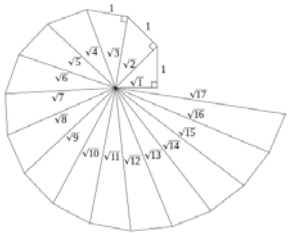 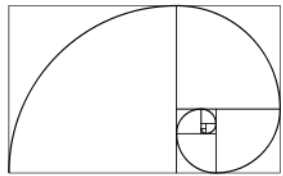 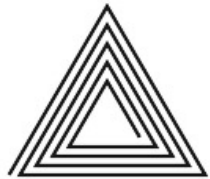 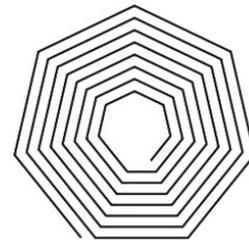

Figure 3: The Theodorus spiral.　　Figure 4: The golden spiral.　　Figure 5: 3-angle spirangle.　　Figure 6: 7-angle spirangle.

Discrete planar and three-dimensional spirals are often used in architecture as spiral staircases.

**3. Spirals with a monotonic curvature function**

The concept of monotonic curvature pieces plays an important role in shape control of curves. Fair curves are used to satisfy highly aesthetic requirements in industrial design and styling, and used for generating aesthetic surfaces like car hoods and roofs. In this section we shortly discuss some fair curves used in CAGD.

**3.1 Pseudospirals**

Pseudospirals are planar spirals whose natural equations can be written in the form [11]

$$\rho = \alpha s^m \Rightarrow \kappa = \frac{1}{\alpha} s^{-m}, (\alpha \in \mathbb{R}, m \in \mathbb{R}), \qquad (1)$$

---

* The name has been given by Annie van Maldeghem, named after a symposium called Matomium, held in 2002 in Brussels.

where $\rho$ is the radius of curvature, $\kappa$ is the curvature and $s$ is the arc length. When $m=1$, this is called the logarithmic spiral, when $m=-1$, the Cornu spiral and when $m=1/2$ it is the involute of a circle. Supposing that the pseudospiral has $\theta=0$ when $s=0$, and integrating in the well-known differential geometry relationship $\frac{\partial \theta}{\partial s} = \rho$, we find the tangent angle at every point of a pseudospiral [19].

$$\theta(s) = \int_0^s \kappa(s) ds = \begin{cases} \alpha \log(s), & m=1 \\ \frac{\alpha}{1-m} s^{1-m}, & otherwise \end{cases}. \qquad (2)$$

Then, the parametric equations of a pseudospiral in terms of natural parameter $s$ can be written as follows:

$$\gamma_{(m=1)}(s) = \begin{pmatrix} \int_0^{\theta(s_1)} \cos(\theta(s)) ds \\ \int_0^{\theta(s_1)} \sin(\theta(s)) ds \end{pmatrix} = \frac{s}{1+\alpha^2} \begin{pmatrix} \cos(\alpha \log(s)) + \alpha \sin(\alpha \log(s)) \\ -\alpha \cos(\alpha \log(s)) + \sin(\alpha \log(s)) \end{pmatrix} \Bigg|_0^{\theta(s_1)}, \qquad (3)$$

and

$$\gamma_{(m \in \mathbb{R} \setminus \{1\})}(s) = \begin{pmatrix} s \, {}_1F_2\left(\frac{1}{2(1-m)}; \frac{1}{2}, \frac{2m-3}{2(m-1)}; -\frac{\alpha^2 s^{2(1-m)}}{4(1-m)^2}\right) \\ \frac{\alpha s^{2-m}}{(m-2)(m-1)} {}_1F_2\left(\frac{m-2}{2(m-1)}; \frac{3}{2}, \frac{3m-4}{2(m-1)}; -\frac{\alpha^2 s^{2(1-m)}}{4(1-m)^2}\right) \end{pmatrix} \Bigg|_0^{\theta(s_1)}, \qquad (4)$$

where upper bound $\theta(s_1)$ for the integrals is computed from Eq. 2 ($s_0=0$) and the hypergeometric function is defined as in [20]

$${}_1F_2(a;b,c;z) = \sum_{n=0}^{\infty} \frac{(a)_n}{(b)_n (c)_n} \frac{z^n}{n!}, \qquad (5)$$

where the formula for a rising factorial is used

$$(a)_n = a(a+1)(a+2)\ldots(a+n-1), (a)_0 = 1. \qquad (6)$$

The pseudospiral can be used as a $G^2$ transition curve between a straight line and any other curve with arbitrary topology.

### 3.2 Log-aesthetic curves

The linear transformation of Eq. 1 leads to the radius of curvature function of so-called log-aesthetic curves (LACs) [6, 7, 8, 9, 10] which are useful for describing visually pleasing shapes. The horizontal axis of the logarithmic curvature graph measures $log\, \rho$ and the vertical axis measures $log(ds/d(log\, \rho)) = log(\rho\, ds/d\rho)$. If the logarithmic curvature graph (LCG) [15] is given by a straight line, a constant $\alpha$ exists such that the following equation is satisfied:

$$log\left(\rho \frac{ds}{d\rho}\right) = \alpha\, log\, \rho + C \qquad (7)$$

where $C$ is a constant. The above equation is called the fundamental equation of aesthetic curves. Rewriting Eq. 7, we obtain

$$\frac{1}{\rho^{\alpha-1}} \frac{ds}{d\rho} = e^C = C_0 \qquad (8)$$

Hence there is some constant $c_0$ such that

$$\rho^{\alpha-1}\frac{d\rho}{ds}=c_0 \quad (9)$$

From the above equation, when $\alpha \neq 0$, the first general equation of aesthetic curves

$$\rho^{\alpha}=c_0 s+c_1 \quad (10)$$

is obtained. If, we obtain the second general equation of aesthetic curves

$$\rho=c_0 e^{c_1 s} \quad (11)$$

the curve which satisfies Eq. 10 or Eq. 11 is called the log-aesthetic curve. The analytic parametric equations of LACs in terms of incomplete Gamma functions were found in [21] and LACs are used to construct a multispiral in [22].

Some applications of LACs are used in the design of car bodies and are shown in Figs. 7 and 8.

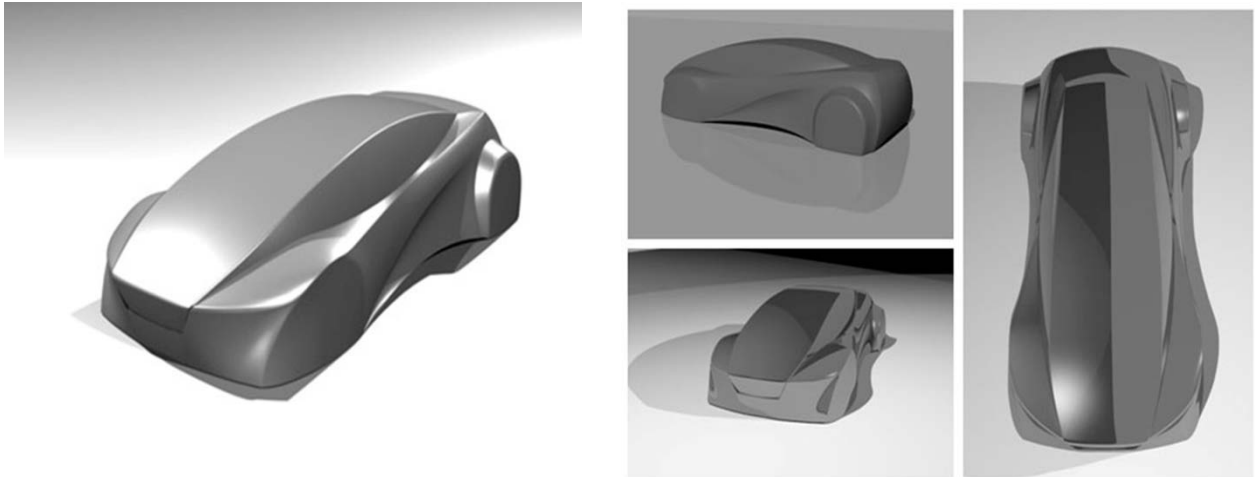

Figure 7: A new conceptual design of a car body using LACs.

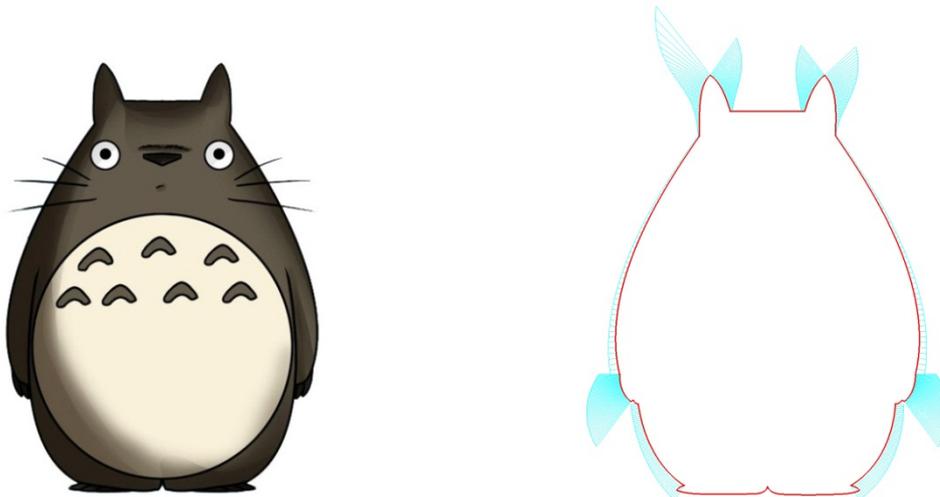

Figure 8: A Totoro (となりのトトロ), character from a Japanese animated fantasy film written and directed by Hayao Miyazaki and produced by Studio Ghibli. Its shape is generated by $G^0$, $G^1$ and $G^2$ log-aesthetic splines. Blue lines show the values of curvature at corresponding points.

### 3.3 Generalized log-aesthetic curves
We can generalize Eq. 10 as follows:

$$\rho = (c_0 s + c_1)^{\frac{1}{\alpha}} + c_2 \quad (12)$$

The right side of the above equation has three parameters, $c_0$, $c_1$ and $c_2$, if we exclude $\alpha$. The meaning of the equation can clearly be interpreted as the radius of curvature is shifted by $c_2$. This type of generalization is called the radius of curvature shift. It is also possible to perform the same extension for the curvature $\kappa$ as follows:

$$\kappa = (c_0 s + c_1)^{-\frac{1}{\alpha}} + c_2 \qquad (13)$$

We define the curve obtained by either of these two types of generalization as the generalized log-aesthetic curves: GLAC [12, 13].

### 3.4 Superspirals

Superspirals [14] are a new and very general family of fair curves, whose radius of curvature is given in terms of a completely monotonic Gauss hypergeometric function defined in Eq. 5. The superspirals are generalizations of LACs and GLACs, as well as other curves whose radius of curvature is a particular case of a completely monotonic Gauss hypergeometric function. They include a huge variety of fair curves with monotonic curvatures and can be computed with a high degree of accuracy using the adaptive Gauss–Kronrod numerical integration method.

In [14] superspirals are defined as a planar curve with a completely monotone radius of curvature given in the form $\rho(\psi) = {}_2F_1(a,b,c,-\psi)$, where $c > b > 0$, $a > 0$. Its corresponding parametric equation in terms of the tangent angle is

$$S(a,b,c,\theta) = \begin{pmatrix} x(\theta) \\ y(\theta) \end{pmatrix} = \begin{pmatrix} \int_0^\theta {}_1F_2(a,b,c,-\psi)\cos(\psi)\mathrm{d}\psi \\ \int_0^\theta {}_1F_2(a,b,c,-\psi)\sin(\psi)\mathrm{d}\psi \end{pmatrix}, 0 \le \theta < +\infty. \qquad (14)$$

They can be used as a transition curve, as it was shown in [14].

### 4. CONCLUSION

In this work we have discussed the variety of planar spirals and their applications in computer-aided design. More attention was paid to the spirals with monotonic curvature function. In addition, the parametric equations of pseudospirals in terms of arclength were expressed in terms of hypergeometric functions, which can be computed with a high degree of accuracy in computer algebra systems (CAS). Our future work will deal with aesthetic surfaces and the algorithms for generating rectangular and triangular aesthetic patches.

УДК 51-74

**О множестве плоских спиралей и их применение в компьютерном дизайне**

[1] Рушан Зиатдинов
[2] Кенджиро Т. Миура

[1] Университет Фатих, Турция
34500 Буюкчекмедже, Стамбул
Кандидат физико-математических наук, Ассистент-Профессор
E-mail: rushanziatdinov@gmail.com, ziatdinov@fatih.edu.tr
[2] Университет Сидзуока, Япония
3-5-1, Джохоку, Нака-ку, Хамаматсу Сидзуока
Доктор технических наук, Профессор
E-mail: tmkmiur@ipc.shizuoka.ac.jp

**Аннотация.** В данной статье рассматривается множество плоских спиральных кривых и их приложения в дизайне объектов реального и виртуального миров. Описываемые семейства кривых с монотонными функциями кривизны общеизвестны в компьютерном геометрическом дизайне как эстетические кривые (кривые высокого качества) и играют важную роль в формообразовании поверхностей высокого качества. Эстетические кривые также используются при решении задач двухточечной Эрмитовой интерполяции с заданными граничными условиями и для построения эстетических сплайнов.

**Ключевые слова:** спираль; псевдоспираль; эстетическая кривая; LAC; GLAC; суперспираль; мультиспираль; монотонность кривизны; кривая высокого качества; эстетический дизайн; гамма-функция; переходная кривая; моделирование фасок; сплайн; компьютерный геометрический дизайн.